\documentclass[amstex,twocolumn,showpacs,floats,floatfix,superscriptaddress,aps,pra]{revtex4}
\usepackage{amssymb}
\usepackage{amsmath}
\usepackage{calc}
\usepackage{graphicx}
\usepackage{bm}

\input{tcilatex}

\begin{document}

\begin{abstract}
The aim of this paper is to analyze the Morris-Shore (MS) transformation in
the case of weak deviation of the condition for equal detunings. For this
generalization of the MS transformation a perturbative solution is derived.
Some elements concerning the general solution of the MS transformation are
also discussed.
\end{abstract}

\pacs{32.80.Bx, 33.80.Be, 03.65.Ge, 42.50.Md}
\author{G. S. Vasilev}
\affiliation{Department of Physics, Sofia University, James Bourchier 5 blvd, 1164 Sofia,
Bulgaria}
\author{N. V. Vitanov}
\affiliation{Department of Physics, Sofia University, James Bourchier 5 blvd, 1164 Sofia,
Bulgaria}
\title{Morris-Shore transformation with unequal detunings}
\date{\today }
\maketitle


\section{Introduction}

The dynamics of a two-state quantum system resulting from the action of a
resonant external pulse of precise temporal area is a classical example in
quantum mechanics. This simple two-state quantum system has an interesting
theoretical extension, from a nondegenerate two-level quantum system (where
only two two quantum states /ground and excited/ are available), to one with
a degenerate ground level and a degenerate excited level. These physical
systems can be found at isolated atoms and molecules, where the states of
well defined angular momentum J, due to the rotational symmetry produces a
degeneracy of 2J+1 magnetic sublevels. For correct description of
laser-induced transitions between quantum states with defined angular
momentum, one must also incorporate several magnetic sublevels( labeled by
M), that may serve as a possible initial states. Hence each of these
magnetic sublevels have a laser-driven excitation dynamic into to set of
excited magnetic sublevels. Other source of multilevel systems can be the
systems, whose mathematical description uses the usual rotating-wave
approximation (RWA). Within the RWA the degeneracies can also be found for
more general multistate quantum systems.

To handle such multilevel systems, in 1983 Morris and Shore have introduced
a coordinate transformation with the general property that could factorize
the dynamic of any two degenerate sets of quantum states, to an equivalent
description involving only independent uncoupled pairs of equations.
Generally, the Morris-Shore (MS) transformation is able to reduces the
coherent quantum dynamics of a coupled degenerate two-level system to a set
of independent nondegenerate two-state systems and a set of uncoupled
states. The mathematical description of MS transformation only requires
deriving of eigenvalues and the eigenstates of a hermitean matrix, which is
a product of interaction matrices. The eigenstates of this hermitean matrix
are the MS states, that represent the independent two-state systems and the
dark states. The eigenvalues represent the MS interactions in the
independent MS two-state systems. The MS transformation is quite general,
but requires that all initial interactions are constant, or share the same
time dependence, and also that all interactions are resonant, or equally
detuned from the upper states. The condition of equally detuned set of
states implies that the lower set of states is degenerate in RWA sense, and
so is the upper set of states.

Having explain the constrains of MS transformation, it is natural to ask
whether they can be relaxed. It turns out that within the framework of weak
deviation of equally detuned set of statrs, perturbative solution can be
derived.

This paper is organized as follows. In Sec. II we review the two-level MS
transformation and in details define the problem for its extension. In Sec.
III we describe the perturbative solution to the problem of generalized MS
transformation. Section IV presents an examples for the generalized MS
transformation . Finally, Sec. V presents a summary of the results.


\section{Definition of the problem\label{Sec-definition}}

We consider a two-level quantum system in the general case when both levels
are degenerate. The ground $\mathbf{a}$ level consists of $N_{a}$ degenerate
states, the excited $\mathbf{b}$ level -- of $N_{b}$ degenerate states and
we assume, without any loss of generality, that $N_{a}\geq N_{b}.$ Every
ground state $\left\vert \psi _{i}\right\rangle $ $(i=1,2,...,N_{a})$ is
coupled via all excited states $\left\vert \psi _{j}\right\rangle $ $%
(j=N_{a}+1,N_{a}+2,...,N_{a}+N_{b})$ with pulsed interactions $\Omega
_{ij}(t)$, each pair of which are on two-photon resonance, as depicted in %
\ref{Fig1}. The excited level may be off single photon resonance by some
detuning $\Delta (t)$ which should be the same for all couplings. We assume
that the couplings share same time dependence $f(t)$
\begin{equation*}
\Omega _{ij}(t)=\chi _{ij}e^{-i\xi t}f(t),
\end{equation*}%
but we allow for different amplitudes $\chi _{ij}$ and thereby pulse areas .

The dynamics of the system is governed by the time-dependent Schr\"{o}dinger
equation which in the usual rotating-wave approximation (RWA) has the form%
\begin{equation}
i\hbar \frac{d}{dt}\mathbf{C}(t)=\mathbf{H}(t)\mathbf{C}(t),  \label{Sch}
\end{equation}%
where the elements of the $\left( N_{a}+N_{b}\right) $-dimensional state
vector $\mathbf{C}(t)$ are the probability amplitudes of the states and the
Hamiltonian in a block-matrix form is given by%
\begin{equation}
\mathbf{H}_{0}(t)=%
\begin{bmatrix}
\mathbf{0}_{N_{a}\times N_{a}} & \mathbf{V}(t)_{N_{a}\times N_{b}} \\
\mathbf{V}^{\dag }(t)_{N_{b}\times N_{a}} & \mathbf{\Delta }(t)_{N_{b}\times
N_{b}}%
\end{bmatrix}%
.  \label{H}
\end{equation}%
Here $\mathbf{0}$ is the $N_{a}$-dimensional square zero matrix, in which
the zero off-diagonal elements indicate the absence of couplings between the
$\mathbf{a}$ states, while the zero diagonal elements show that the $\mathbf{%
a}$ states have the same energy, which is taken as the zero of the energy
scale. The matrix $\mathbf{\Delta }(t)$ is an $N_{b}$-dimensional square
diagonal matrix, with $\delta (t)$ on the diagonal, $\mathbf{\Delta }(t)$ $=$
$\delta (t)\mathbf{1}_{N_{b}}$ In Eq.(\ref{H}) $\mathbf{V}(t)$ is an $%
(N_{a}\times N_{b})$-dimensional matrix, which elements are the couplings
between the ground and the excited states, $\mathbf{V}^{\dag }(t)$ is its
hermitian conjugate

\begin{eqnarray}
\mathbf{V}(t) &=&%
\begin{bmatrix}
\Omega _{11} & \Omega _{12} & \cdots & \Omega _{1N_{b}} \\
\Omega _{21} & \Omega _{22} & \cdots & \Omega _{2N_{b}} \\
\cdots & \cdots & \ddots & \vdots \\
\Omega _{N_{a}1} & \Omega _{N_{a}2} & \cdots & \Omega _{N_{a}N_{b}}%
\end{bmatrix}
\label{V} \\
&=&[\left\vert \Omega _{1}\right\rangle ,\left\vert \Omega _{2}\right\rangle
,...,\left\vert \Omega _{N_{b}}\right\rangle ],  \label{V-vectors}
\end{eqnarray}%
where $\left\vert \Omega _{n}\right\rangle $ $(n=1,2,...,N_{b})$ are $N_{b}$%
-dimensional vectors, composed of the interaction energies of the $n$th
state of the $\mathbf{b}$ set with all states of the $\mathbf{a}$ set,%
\begin{equation}
\left\vert \Omega _{n}\right\rangle =%
\begin{bmatrix}
\Omega _{1n} \\
\Omega _{2n} \\
\vdots \\
\Omega _{N_{a}n}%
\end{bmatrix}%
.  \label{O}
\end{equation}

If the Hamiltonian describes the atom-laser interaction, in the presence of
weak magnetic field the degeneracy of the states is lifted. Due to the
external field we have an extra terms, which can be assumed as a
perturbation to the exact MS solution. In this case we have $\mathbf{H}(t)=%
\mathbf{H}_{0}(t)+\mathbf{D}(t),$where $\mathbf{D}(t)$ is diagonal matrix
with dimension $N_{a}+N_{b}$ and generally $\mathbf{D}(t)$ $\neq $ $d(t)%
\mathbf{1}_{N_{b}}$. The aim of this paper is to analyze the MS
transformation in the case of weak deviation of the $\mathbf{\Delta }(t)$ $=
$ $\delta (t)\mathbf{1}_{N_{b}}$ condition, which in necessary for
factorization towards set of two-state systems and set of decouple states.
Some elements concerning the general solution of the MS transformation will
be also discussed.

We will start our discussion by a brief description of the exact
Morris-Shore (MS) transformation.

\section{Exact analytical solution}

\subsection{Morris-Shore transformation}

\bigskip

Morris and Shore \cite{Morris83} have shown that any degenerate two-level
system, in which all couplings share the same time dependence and the same
detuning, can be reduced with a constant unitary transformation $\mathbf{S}$
to an equivalent system comprising only independent two-state systems and
uncoupled (dark) states, as shown in Fig. \ref{Fig-system}. This
time-independent transformation is given by
\begin{equation}
|\psi _{i}\rangle =\sum_{k}S_{ik}^{\ast }|\widetilde{\psi }_{k}\rangle \quad
\Longleftrightarrow \quad |\widetilde{\psi }_{k}\rangle =\sum_{i}S_{ik}|\psi
_{i}\rangle ,  \label{transformation}
\end{equation}%
where the tildas denote the MS basis hereafter. The constant transformation
matrix $\mathbf{S}$ can be represented in the block-matrix form
\begin{equation}
\mathbf{S}=\left[
\begin{array}{cc}
\mathbf{A} & \mathbf{O} \\
\mathbf{O} & \mathbf{B}%
\end{array}%
\right] ,  \label{W}
\end{equation}%
where $\mathbf{A}$ is a unitary $N_{a}$-dimensional square matrix and $%
\mathbf{B}$ is a unitary $N_{b}$-dimensional square matrix, $\mathbf{AA}%
^{\dagger }=\mathbf{A}^{\dagger }\mathbf{A}=\mathbf{1}_{N_{a}}$ and $\mathbf{%
BB}^{\dagger }=\mathbf{B}^{\dagger }\mathbf{B}=\mathbf{1}_{N_{b}}$. The
constant matrices $\mathbf{A}$ and $\mathbf{B}$ mix only sublevels of a
given level: $\mathbf{A}$ mixes the $a$ sublevels and $\mathbf{B}$ mixes the
$b$ sublevels. The transformed MS Hamiltonian has the form%
\begin{equation}
\widetilde{\mathbf{H}}(t)=\mathbf{SH}(t)\mathbf{S}^{\dagger }=\left[
\begin{array}{cc}
\mathbf{0} & \widetilde{\mathbf{V}} \\
\widetilde{\mathbf{V}}^{\dagger } & \mathbf{\Delta }(t)%
\end{array}%
\right] ,  \label{H-MS}
\end{equation}%
where
\begin{equation}
\widetilde{\mathbf{V}}=\mathbf{AVB}^{\dagger }.  \label{M}
\end{equation}%
The $N_{a}\times N_{b}$ matrix $\widetilde{\mathbf{V}}$ may have $%
N_{d}=N_{a}-N_{b}$ null rows (if $N_{a}>N_{b}$), which correspond to
decoupled states. The decomposition of $\mathbf{H}$ into a set of
independent two-state systems requires that, after removing the null rows, $%
\widetilde{\mathbf{V}}$ reduces to a $N_{b}$-dimensional diagonal matrix;
let us denote its diagonal elements by $\lambda _{n}$ ($n=1,2,\ldots ,N_{b}$%
).

It follows from Eq. (\ref{M}) that
\begin{subequations}
\label{MM}
\begin{eqnarray}
\widetilde{\mathbf{V}}\widetilde{\mathbf{V}}^{\dagger } &=&\mathbf{AVV}%
^{\dagger }\mathbf{A}^{\dagger },  \label{MM+} \\
\widetilde{\mathbf{V}}^{\dagger }\widetilde{\mathbf{V}} &=&\mathbf{BV}%
^{\dagger }\mathbf{VB}^{\dagger }.  \label{M+M}
\end{eqnarray}%
Hence $\mathbf{A}$ and $\mathbf{B}$ are defined by the condition that they
diagonalize $\mathbf{VV}^{\dagger }$ and $\mathbf{V}^{\dagger }\mathbf{V}$,
respectively. It is important to note that the square matrices $\mathbf{VV}%
^{\dagger }$ and $\mathbf{V}^{\dagger }\mathbf{V}$ have different
dimensions, $N_{a}$ and $N_{b}$, respectively. Because all elements of $%
\mathbf{V}$ are constant, $\mathbf{A}$ and $\mathbf{B}$ are also constant;
the elements $\lambda _{n}$ are constant too. It is straightforward to show
that the $N_{b}$ eigenvalues of $\mathbf{V}^{\dagger }\mathbf{V}$ are all
non-negative; according to Eqs. (\ref{M}) and (\ref{MM}) they are $\lambda
_{n} ^{2}$ ($n=1,2,\ldots ,N_{b}$). The matrix $\mathbf{VV}^{\dagger }$ has
the same eigenvalues and additional $N_{d}=N_{a}-N_{b}$ zero eigenvalues.

The MS Hamiltonian (\ref{H-MS}) has the explicit form
\end{subequations}
\begin{equation}
\widetilde{\mathbf{H}}=\left[
\begin{tabular}{cc}
$\mathbf{0}_{N_{d}}$ & $\mathbf{0}$ \\
$\mathbf{0}$ & $%
\begin{array}{cccccccc}
0 & 0 & \cdots & 0 & \lambda _{1} & 0 & \cdots & 0 \\
0 & 0 & \cdots & 0 & 0 & \lambda _{2} & \cdots & 0 \\
\vdots & \vdots & \ddots & \vdots & \vdots & \vdots & \ddots & \vdots \\
0 & 0 & \cdots & 0 & 0 & 0 & \cdots & \lambda _{N_{b}} \\
\lambda _{1} & 0 & \cdots & 0 & \Delta & 0 & \cdots & 0 \\
0 & \lambda _{2} & \cdots & 0 & 0 & \Delta & \cdots & 0 \\
\vdots & \vdots & \ddots & \vdots & \vdots & \vdots & \ddots & \vdots \\
0 & 0 & \cdots & \lambda _{N_{b}} & 0 & 0 & \cdots & \Delta%
\end{array}%
$%
\end{tabular}%
\right] .  \label{H-MS explicit}
\end{equation}%
The structure of the MS Hamiltonian (\ref{H-MS explicit}) shows that in the
MS basis the dynamics is decomposed into sets of $N_{d}$ decoupled single
states, and $N_{b}$ independent two-state systems $|\widetilde{\psi }%
_{n}^{a}\rangle \leftrightarrow |\widetilde{\psi }_{n}^{b}\rangle $ ($%
n=1,2,\ldots ,N_{b}$), each composed of an $a$ state $|\widetilde{\psi }%
_{n}^{a}\rangle $ and a $b$ state $|\widetilde{\psi }_{n}^{b}\rangle $, and
driven by the RWA Hamiltonians,
\begin{equation}
\widetilde{\mathbf{H}}_{n}(t)=\left[
\begin{array}{cc}
0 & \lambda _{n} \\
\lambda _{n} & \Delta (t)%
\end{array}%
\right] \quad (n=1,2,\ldots ,N_{b}).  \label{Hn}
\end{equation}%
Each of these two-state Hamiltonians has the same detuning $\Delta (t)$, but
they differ in the couplings $\lambda _{n}$. Each of the new $a$ states $|%
\widetilde{\psi }_{n}^{a}\rangle $ is the eigenstate of $\mathbf{VV}%
^{\dagger }$ corresponding to the eigenvalue $\lambda _{n}^{2}$, whereas
each of the new $b$ states $|\widetilde{\psi }_{n}^{b}\rangle $ is the
eigenstate of $\mathbf{V}^{\dagger }\mathbf{V}$, corresponding to the same
eigenvalue $\lambda _{n}^{2}$. The square root of this common eigenvalue, $%
\lambda _{n}$, represents the coupling in the respective independent MS
two-state system $|\widetilde{\psi }_{n}^{a}\rangle \leftrightarrow |%
\widetilde{\psi }_{n}^{b}\rangle $ ($n=1,2,\ldots ,N_{b}$). The $N_{d}$ zero
eigenvalues of $\mathbf{VV}^{\dagger }$ correspond to decoupled (dark)
states in the $a$ set (we assume thoughout that $N_{a}\geqq N_{b}$;
therefore, dark states, if any, are in the $a$ set). The dark states are
decoupled from the dynamical evolution because they are driven by
one-dimensional null Hamiltonians.

\begin{figure}[tb]
\includegraphics[width=75mm]{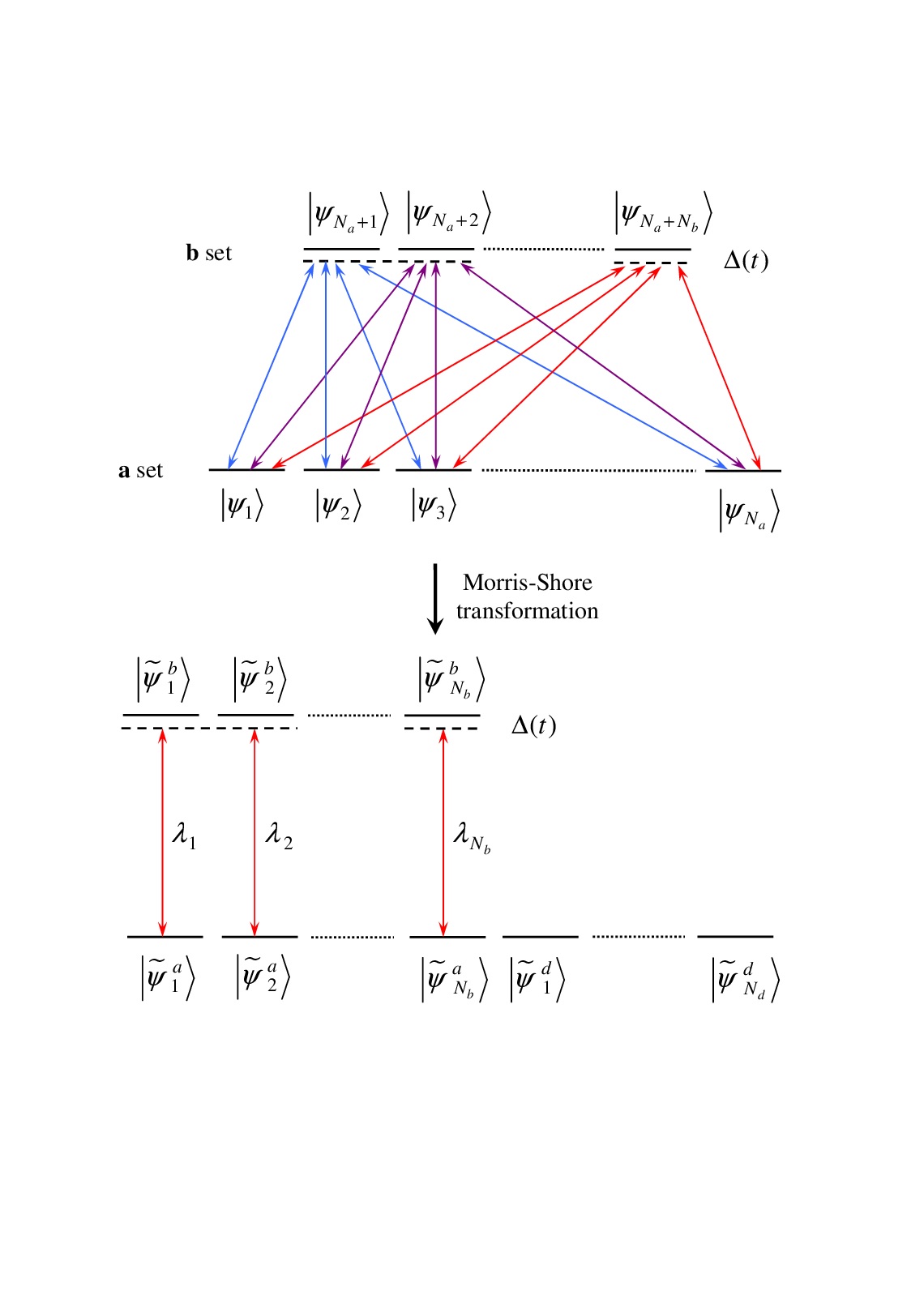}
\caption{Scheme of the Morris-Shore transformation, where a
multistate system consisting of two coupled sets of degenerate
levels is decomposed into a set of independent nondegenerate
two-state systems and a set of decoupled states.} \label{MS}
\end{figure}

The MS decomposition described above allows us to reduce the
degenerate two-level problem to a set of $N_{d}$ nondegenerate
two-state problems, wherein the detuning is unchanged while the
couplings $\lambda _{n}$ are combinations of the initial couplings
between the $a$ and $b$ states and
defined as the square roots of the eigenvalues of $\mathbf{V}^{\dagger }%
\mathbf{V}$.

From the vector form (\ref{V-vectors}) of $\mathbf{V}$ we obtain
\begin{subequations}
\begin{eqnarray}
\mathbf{VV}^{\dagger } &=&\sum_{n=1}^{N_{b}}\left\vert \Omega
_{n}\right\rangle \left\langle \Omega _{n}\right\vert ,  \label{VV+} \\
\mathbf{V}^{\dagger }\mathbf{V} &=&\left[
\begin{array}{cccc}
\left\langle \Omega _{1}|\Omega _{1}\right\rangle & \left\langle \Omega
_{1}|\Omega _{2}\right\rangle & \cdots & \left\langle \Omega _{1}|\Omega
_{N_{b}}\right\rangle \\
\left\langle \Omega _{2}|\Omega _{1}\right\rangle & \left\langle \Omega
_{2}|\Omega _{2}\right\rangle & \cdots & \left\langle \Omega _{2}|\Omega
_{N_{b}}\right\rangle \\
\vdots & \vdots & \ddots & \vdots \\
\left\langle \Omega _{N_{b}}|\Omega _{1}\right\rangle & \left\langle \Omega
_{N_{b}}|\Omega _{2}\right\rangle & \cdots & \left\langle \Omega
_{N_{b}}|\Omega _{N_{b}}\right\rangle%
\end{array}%
\right] .  \label{V+V}
\end{eqnarray}%
Note that $\mathbf{V}^{\dagger }\mathbf{V}$ is the Gramm matrix for the set
of vectors $\left\{ \left\vert \Omega _{n}\right\rangle \right\}
_{n=1}^{N_{b}}$. Thus if all these vectors are linearly independent then $%
\det \mathbf{V}^{\dagger }\mathbf{V}\neq 0$ and all eigenvalues of $\mathbf{V%
}^{\dagger }\mathbf{V}$ are nonzero \cite{Gantmacher}; however, this
assumption is unnecessary.

We assume that we can find the eigenvalues $\lambda _{n}^{2}$ $(n=1,2,\ldots
,N_{b})$ of the matrices (\ref{VV+}) and (\ref{V+V}), and the corresponding
orthonormalized eigenvectors: the $N_{b}$ coupled eigenstates $|\widetilde{%
\psi }_{n}^{a}\rangle $ of $\mathbf{VV}^{\dagger }$ and $|\widetilde{\psi }%
_{n}^{b}\rangle $ of $\mathbf{V}^{\dagger }\mathbf{V}$, and the $N_{d}$
decoupled eigenstates $|\widetilde{\psi }_{k}^{d}\rangle $ of $\mathbf{VV}%
^{\dagger }$. We use these eigenstates to construct the transformation
matrices as
\end{subequations}
\begin{equation}
\mathbf{A}=\left[
\begin{array}{c}
\langle \widetilde{\psi }_{1}^{d}| \\
\vdots \\
\langle \widetilde{\psi }_{N_{d}}^{d}| \\
\langle \widetilde{\psi }_{1}^{a}| \\
\vdots \\
\langle \widetilde{\psi }_{N_{b}}^{a}|%
\end{array}%
\right] ,\quad \mathbf{B}=\left[
\begin{array}{c}
\langle \widetilde{\psi }_{1}^{b}| \\
\vdots \\
\langle \widetilde{\psi }_{N_{b}}^{b}|%
\end{array}%
\right] .  \label{A B}
\end{equation}

\subsection{The MS\ propagators}

The propagator the independent MS\ two-state systems $|\widetilde{\psi }%
_{n}^{a}\rangle \leftrightarrow |\widetilde{\psi }_{n}^{b}\rangle $ ($%
n=1,2,\ldots ,N_{b}$), defined by

\begin{equation}
\left[
\begin{array}{c}
\widetilde{C}_{n}^{a}(t_{f}) \\
\widetilde{C}_{n}^{b}(t_{f})%
\end{array}%
\right] =\widetilde{\mathbf{U}}_{n}(t_{f},t_{i})\left[
\begin{array}{c}
\widetilde{C}_{n}^{a}(t_{i}) \\
\widetilde{C}_{n}^{b}(t_{i})%
\end{array}%
\right] ,
\end{equation}%
is unitary and can be parameterized in terms of the Cayley-Klein parameters%
\begin{equation}
\widetilde{\mathbf{U}}_{n}(t_{f},t_{i})=%
\begin{bmatrix}
\alpha _{n} & -\beta _{n}^{\ast } \\
\beta _{n} & \alpha _{n}^{\ast }%
\end{bmatrix}%
.  \label{Un}
\end{equation}%
Here, $\alpha _{n}$ and $\beta _{n}$ are the Cayley-Klein parameters and
they obey the relation%
\begin{equation}
\left\vert \alpha \right\vert ^{2}+\left\vert \beta \right\vert ^{2}=1.
\end{equation}

\subsection{The propagator in the diabatic basis}

By taking into account the MS propagators (\ref{Un}) for the two-state MS
systems, the ordering of the states, and the MS Hamiltonian (\ref{H-MS
explicit}), the propagator of the entire system in the MS basis is written as%
\begin{equation}
\widetilde{\mathbf{U}}=\left[
\begin{array}{cc}
\mathbf{1}_{N_{d}} & \mathbf{0} \\
\mathbf{0} &
\begin{array}{cccccccc}
\alpha _{1} & 0 & \cdots & 0 & -\beta _{1}^{\ast } & 0 & \cdots & 0 \\
0 & \alpha _{2} & \cdots & 0 & 0 & -\beta _{2}^{\ast } & \cdots & 0 \\
\vdots & \vdots & \ddots & \vdots & \vdots & \vdots & \ddots & \vdots \\
0 & 0 & \cdots & \alpha _{N_{b}} & 0 & 0 & \cdots & -\beta _{N_{b}}^{\ast }
\\
\beta _{1} & 0 & \cdots & 0 & \alpha _{1}^{\ast } & 0 & \cdots & 0 \\
0 & \beta _{2} & \cdots & 0 & 0 & \alpha _{2}^{\ast } & \cdots & 0 \\
\vdots & \vdots & \ddots & \vdots & \vdots & \vdots & \ddots & \vdots \\
0 & 0 & \cdots & \beta _{N_{b}} & 0 & 0 & \cdots & \alpha _{N_{b}}^{\ast }%
\end{array}%
\end{array}%
\right] .  \label{U-MS}
\end{equation}%
By taking into account the completeness relations
\begin{subequations}
\label{completeness}
\begin{gather}
\sum_{n=1}^{N_{b}}|\widetilde{\psi }_{n}^{a}\rangle \langle \widetilde{\psi }%
_{n}^{a}|+\sum_{k=1}^{N_{d}}|\widetilde{\psi }_{k}^{d}\rangle \langle
\widetilde{\psi }_{k}^{d}|=\mathbf{1}_{N_{a}},  \label{completeness a} \\
\sum_{n=1}^{N_{b}}|\widetilde{\psi }_{n}^{b}\rangle \langle \widetilde{\psi }%
_{n}^{b}|=\mathbf{1}_{N_{b}},  \label{completeness b}
\end{gather}%
it is straightforward to show that the propagator in the original basis $%
\mathbf{U}=\mathbf{S^{\dagger }}\widetilde{\mathbf{U}}\mathbf{S}$ reads
\end{subequations}
\begin{equation}
\mathbf{U}=\left[
\begin{array}{cc}
\mathbf{1}_{N_{a}}+\sum_{n=1}^{N_{b}}(\alpha _{n}-1)|\widetilde{\psi }%
_{n}^{a}\rangle \langle \widetilde{\psi }_{n}^{a}| & -\sum_{n=1}^{N_{b}}%
\beta _{n}^{\ast }|\widetilde{\psi }_{n}^{a}\rangle \langle \widetilde{\psi }%
_{n}^{b}| \\
\sum_{n=1}^{N_{b}}\beta _{n}|\widetilde{\psi }_{n}^{b}\rangle \langle
\widetilde{\psi }_{n}^{a}| & \sum_{n=1}^{N_{b}}\alpha _{n}^{\ast }|%
\widetilde{\psi }_{n}^{b}\rangle \langle \widetilde{\psi }_{n}^{b}|%
\end{array}%
\right] .  \label{U}
\end{equation}%
Note that the propagator does not depend on the decoupled states $|%
\widetilde{\psi }_{k}^{d}\rangle $ $(k=1,2,\ldots ,N_{d})$, which are
excluded by using Eq. (\ref{completeness a}). This has to be expected
because, owing to their degeneracy, the choice of the decoupled states is
not unique, since any superposition of them is also a zero-eigenvalue
eigenstate of $\mathbf{VV}^{\dagger }$. Because the dynamics in the original
basis must not depend on such arbitrariness, the propagator $\mathbf{U}$
must not depend on the decoupled states at all.

\section{MS with nondegenerate(unequal) detunings}

We

\begin{equation}
\mathbf{H}(t)=%
\begin{bmatrix}
\mathbf{D}_{1}(t)_{N_{a}\times N_{a}} & \mathbf{V}(t)_{N_{a}\times N_{b}} \\
\mathbf{V}^{\dag }(t)_{N_{b}\times N_{a}} & \left( \mathbf{\Delta }(t)+%
\mathbf{D}_{2}(t)\right) _{N_{b}\times N_{b}}%
\end{bmatrix}%
=\mathbf{H}_{0}(t)+\mathbf{D}(t).
\end{equation}

where $\mathbf{D}(t)_{(N_{a}+N_{b})\times (N_{a}+N_{b})}$ is diagonal
matrix, given by

\begin{equation*}
\mathbf{D}(t)=%
\begin{bmatrix}
\mathbf{D}_{1}(t)_{N_{a}\times N_{a}} & \mathbf{0} \\
\mathbf{0} & \mathbf{D}_{2}(t)_{N_{b}\times N_{b}}%
\end{bmatrix}%
\end{equation*}

\begin{eqnarray}
\mathbf{D}_{1}(t) &=&\text{diag}\left[
d_{1}^{1}(t),d_{2}^{1}(t),..,d_{N_{a}}^{1}(t)\right] ,  \label{D} \\
\mathbf{D}_{2}(t) &=&\text{diag}\left[
d_{1}^{2}(t),d_{2}^{2}(t),..,d_{N_{b}}^{2}(t)\right] ,
\end{eqnarray}

The main difficulty for finding exact solution is that in the MS basis we
have%
\begin{eqnarray*}
\widetilde{\mathbf{H}}(t) &=&\mathbf{SH}(t)\mathbf{S}^{\dagger }=\left[
\begin{array}{cc}
\mathbf{A\mathbf{D}}_{1}\mathbf{(t)A}^{\dagger } & \widetilde{\mathbf{V}} \\
\widetilde{\mathbf{V}}^{\dagger } & \mathbf{\Delta }(t)+\mathbf{B\mathbf{D}}%
_{2}\mathbf{(t)B}^{\dagger }%
\end{array}%
\right] \\
&=&\mathbf{S}(\mathbf{H}_{0}(t)+\mathbf{D}(t))\mathbf{S}^{\dagger }=\mathbf{H%
}_{0}^{MS}(t)+\mathbf{SD}(t)\mathbf{S}^{\dagger }
\end{eqnarray*}

The matrices $\mathbf{A\mathbf{D}}_{1}\mathbf{(t)A}^{\dagger }\neq $ $%
\mathbf{\mathbf{D}}_{1}\mathbf{(t)}$ and $\mathbf{B\mathbf{D}}_{2}\mathbf{%
(t)B}^{\dagger }\neq $ $\mathbf{\mathbf{D}}_{2}\mathbf{(t)}$ are not
diagonal This simply means in the MS basis the dynamics cannot be decomposed
into sets of $N_{d}$ decoupled single states, and $N_{b}$ independent
two-state systems. Hereafter for convenience we will assume $\mathbf{D}%
_{1}(t)=0$ and the notation $\mathbf{\mathbf{D}}_{2}\mathbf{(t)=D}(t)$ will
be used.

Instead of exact factorization we look at solution that can be expressed in
perturbative form. The Hamiltonian in a block-matrix form is given by $%
\mathbf{H}(t)=\mathbf{H}_{0}(t)+\epsilon \mathbf{D}(t).$ We will look for
perturbative solution for the MS problem and for this reason we will have
the following definition for the transformation matrix $\mathbf{S.}$

\begin{equation}
\mathbf{S}=\mathbf{S}_{0}+\varepsilon \mathbf{S}_{1}+\varepsilon ^{2}\mathbf{%
S}_{2}+...  \label{S-series}
\end{equation}

where the additional condition $\mathbf{\dot{S}=0}$ is imposed. This simply
means that all terms from the series expansion in Eq.(\ref{S-series}) are
constant matrices. Also in reason to simplify the calculations and without
loss of generality the $\mathbf{S}_{i}$ matrices have the block-diagonal
form, similar to (\ref{W})%
\begin{equation}
\mathbf{S}_{i}=\left[
\begin{array}{cc}
\mathbf{A}_{i} & \mathbf{0} \\
\mathbf{0} & \mathbf{B}_{i}%
\end{array}%
\right] .  \label{S-matrix-form}
\end{equation}

According to the Eq.(\ref{transformation}) and using the series expansion
for the transformation matrix $\mathbf{S}$ given in Eq.(\ref{S-series}) we
will obtain
\begin{eqnarray}
\mathbf{C}(t) &=&\mathbf{S}^{\dagger }\mathbf{\tilde{C}}(t)  \label{C-eps} \\
&=&\left( \mathbf{S}_{0}+\varepsilon \mathbf{S}_{1}+\varepsilon ^{2}\mathbf{S%
}_{2}+...\right) \mathbf{\tilde{C}}(t)  \notag \\
&=&\mathbf{\tilde{C}}_{0}(t)+\varepsilon \mathbf{\tilde{C}}%
_{1}(t)+\varepsilon ^{2}\mathbf{\tilde{C}}_{2}(t)+...  \notag
\end{eqnarray}

From the expression above is clear the meaning of the $\mathbf{\tilde{C}}%
_{i}(t)$ state vectors.

Using the series expansion for the transformation matrix $\mathbf{S}$ we
will obtain the following expression for the Hamiltonian in the new basis%
\begin{eqnarray}
\widetilde{\mathbf{H}}(t) &=&\mathbf{SH}(t)\mathbf{S}^{\dagger }=
\label{H-MSbasis-eps} \\
&=&\left( \sum_{i=0}\varepsilon ^{i}\mathbf{S}_{i}\right) \left[ \mathbf{H}%
_{0}(t)+\varepsilon \mathbf{D}(t)\right] \left( \sum_{i=0}\varepsilon ^{i}%
\mathbf{S}_{i}^{\dagger }\right) \\
&=&\mathbf{S}_{0}\mathbf{H}_{0}\mathbf{S}_{0}^{\dagger }+\varepsilon \left[
\mathbf{S}_{0}\mathbf{DS}_{0}^{\dagger }+\mathbf{S}_{1}\mathbf{H}_{0}\mathbf{%
S}_{0}^{\dagger }+\mathbf{S}_{0}\mathbf{H}_{0}\mathbf{S}_{1}^{\dagger }%
\right]  \notag \\
&&+\varepsilon ^{2}\left[ \mathbf{S}_{0}\mathbf{DS}_{1}^{\dagger }+\mathbf{S}%
_{1}\mathbf{DS}_{0}^{\dagger }+\right.  \notag \\
&&\left. \mathbf{S}_{0}\mathbf{H}_{0}\mathbf{S}_{2}^{\dagger }+\mathbf{S}_{1}%
\mathbf{H}_{0}\mathbf{S}_{1}^{\dagger }+\mathbf{S}_{2}\mathbf{H}_{0}\mathbf{S%
}_{2}^{\dagger }\right] +O(\varepsilon ^{3})  \notag
\end{eqnarray}

In the above expression for the Hamiltonian only the first two terms in the
power expansion of $\varepsilon $ are given, but the reader can easily take
more terms for this perturbative expansion.

Let the matrix $\mathbf{S}_{0}$ be chosen such that
\begin{equation}
\mathbf{S}_{0}\mathbf{H}_{0}\mathbf{S}_{0}^{\dagger }=\mathbf{H}_{0}^{MS}(t),%
\text{ }\mathbf{S}_{0}\text{- definition}  \label{S0-def}
\end{equation}%
and $\mathbf{H}_{0}^{MS}(t)$ has the form

\begin{equation*}
\mathbf{H}_{0}^{MS}(t)=\left[
\begin{array}{cc}
\mathbf{0} & \widetilde{\mathbf{V}} \\
\widetilde{\mathbf{V}}^{\dagger } & \mathbf{\Delta }%
\end{array}%
\right] ,
\end{equation*}%
where $\widetilde{\mathbf{V}}$ is given by

\begin{equation*}
\widetilde{\mathbf{V}}=\left[
\begin{array}{c}
\mathbf{0}_{N_{d}\times N_{b}} \\
\mathbf{\Lambda }_{N_{b}\times N_{b}}%
\end{array}%
\right] .
\end{equation*}

The square matrix $\mathbf{\Lambda }$ is diagonal,
\begin{equation*}
\mathbf{\Lambda }(t)=\text{diag}\left[ \lambda _{1}(t),\lambda
_{2}(t),..,\lambda _{N_{b}}(t)\right]
\end{equation*}

This choice for the matrix $\mathbf{S}_{0}$ corresponds to neglecting the
perturbative term $\varepsilon \mathbf{D}(t).$

Having in mind that the matrix $\mathbf{S}_{0}$ is computed, using Eq.(\ref%
{H-MSbasis-eps}) we can continue and take the corrections up to first order
of $\varepsilon $ by choosing matrix $\mathbf{S}_{1}$ from the equation

\begin{equation}
\mathbf{S}_{1}\mathbf{H}_{0}(t)\mathbf{S}_{0}^{\dagger }+\mathbf{S}_{0}%
\mathbf{H}_{0}(t)\mathbf{S}_{1}^{\dagger }+\mathbf{S}_{0}\mathbf{D}(t)%
\mathbf{S}_{0}^{\dagger }=0,\text{ }\mathbf{S}_{1}\text{- definition}
\label{S1-def}
\end{equation}

Using the block-matrix forms for $\mathbf{H}_{0}(t),$ $\mathbf{D}(t),$ $%
\mathbf{S}_{0}$ and $\mathbf{S}_{1}$ matrices and writing the third term to
the rhs of the equation the above expression reads
\begin{widetext}
\begin{equation}
\left[
\begin{array}{cc}
\mathbf{0} & \mathbf{A}_{1}\mathbf{\mathbf{V}B}_{0}^{\dagger } \\
\mathbf{B}_{1}\mathbf{\mathbf{V}}^{\dagger }\mathbf{A}_{0}^{\dagger } &
\mathbf{B}_{1}\mathbf{\Delta B}_{0}^{\dagger }%
\end{array}%
\right] +\left[
\begin{array}{cc}
\mathbf{0} & \mathbf{A}_{0}\mathbf{\mathbf{V}B}_{1}^{\dagger } \\
\mathbf{B}_{0}\mathbf{\mathbf{V}}^{\dagger }\mathbf{A}_{1}^{\dagger } &
\mathbf{B}_{0}\mathbf{\Delta B}_{1}^{\dagger }%
\end{array}%
\right] =-\left[
\begin{array}{cc}
\mathbf{0} & \mathbf{0} \\
\mathbf{0} & \mathbf{B}_{0}\mathbf{DB}_{1}^{\dagger }%
\end{array}%
\right]  \label{S1-matrix-form}
\end{equation}
\end{widetext}
The above block-matrix equation is equivalent to the system of matrix
equations%
\begin{eqnarray}
\mathbf{A}_{1}\mathbf{\mathbf{V}B}_{0}^{\dagger }+\mathbf{A}_{0}\mathbf{%
\mathbf{V}B}_{1}^{\dagger } &=&0  \label{S1-matrix-system} \\
\mathbf{B}_{1}\mathbf{\mathbf{V}}^{\dagger }\mathbf{A}_{0}^{\dagger }+%
\mathbf{B}_{0}\mathbf{\mathbf{V}}^{\dagger }\mathbf{A}_{1}^{\dagger } &=&0
\label{S1-b} \\
\mathbf{B}_{1}\mathbf{\Delta B}_{0}^{\dagger }+\mathbf{B}_{0}\mathbf{\Delta B%
}_{1}^{\dagger } &=&-\mathbf{B}_{0}\mathbf{DB}_{1}^{\dagger }  \label{S1-c}
\end{eqnarray}

The first and the second equation from the above system are equivalent under
the hermitian conjugate operation. We want to note that from Eq.(\ref{S0-def}%
) the $\mathbf{A}_{0}$ and $\mathbf{B}_{0}$ matrices are known. Hence the
system of equation given by Eq.(\ref{S1-b}) and Eq.(\ref{S1-c}), can be
solved for $\mathbf{A}_{1}$ and $\mathbf{B}_{1}$ matrices, which defines the
transformation matrix $\mathbf{S}_{1}.$

Before we continue with $\mathbf{S}_{2}$ matrix computation we would like to
simplify the matrix equation Eq.(\ref{S1-c}). Remember that the matrix $%
\mathbf{\Delta }(t)$ $=$ $\delta (t)\mathbf{1}_{N_{b}}$, one can use the
commutation relations $\mathbf{B}_{1}\mathbf{\Delta =\Delta B}_{1}$ and $%
\mathbf{B}_{0}\mathbf{\Delta =\Delta B}_{0}$, and the unitary properties $%
\mathbf{B}_{0}$ and $\mathbf{B}_{1}$\ to transform Eq.(\ref{S1-c}).

\begin{eqnarray*}
\mathbf{\Delta }\left( \mathbf{B}_{1}\mathbf{B}_{0}^{\dagger }+\mathbf{B}_{0}%
\mathbf{B}_{1}^{\dagger }\right) &=&-\mathbf{B}_{0}\mathbf{DB}_{1}^{\dagger }
\\
\left( \mathbf{B}_{1}\mathbf{B}_{0}^{\dagger }+\mathbf{B}_{0}\mathbf{B}%
_{1}^{\dagger }\right) &=&-\mathbf{B}_{0}\mathbf{\Delta }^{-1}\mathbf{DB}%
_{1}^{\dagger }
\end{eqnarray*}

Using the unitary properties for the matrices $\mathbf{B}_{0}$ and $\mathbf{B%
}_{1}$\ for Eq.(\ref{S1-c}) the system of matrix equation,\ which defines
the transformation matrix $\mathbf{S}_{1}$ can be written in the form

\begin{eqnarray}
\mathbf{A}_{1}\mathbf{\mathbf{V}B}_{0}^{\dagger }+\mathbf{A}_{0}\mathbf{%
\mathbf{V}B}_{1}^{\dagger } &=&0  \label{S1-system} \\
\mathbf{B}_{0}^{\dagger }\mathbf{B}_{1}+\mathbf{B}_{1}^{\dagger }\mathbf{B}%
_{0} &=&-\mathbf{\Delta }^{-1}\mathbf{D}  \notag
\end{eqnarray}

The algorithm for deriving expression for the $\mathbf{S}_{2}$ matrix is
similar to this for $\mathbf{S}_{1}$ matrix. Having in mind that the
matrices $\mathbf{S}_{0}$ and $\mathbf{S}_{1}$ are computed, we can continue
and take the corrections up to $\varepsilon ^{2}$ by choosing matrix $%
\mathbf{S}_{2}$ from the equation

\begin{eqnarray}
&&\mathbf{S}_{0}\mathbf{D}(t)\mathbf{S}_{1}^{\dagger }+\mathbf{S}_{1}\mathbf{%
D}(t)\mathbf{S}_{0}^{\dagger }+  \label{S2} \\
&&\mathbf{S}_{0}\mathbf{H}_{0}(t)\mathbf{S}_{2}^{\dagger }+\mathbf{S}_{1}%
\mathbf{H}_{0}(t)\mathbf{S}_{1}^{\dagger }+\mathbf{S}_{2}\mathbf{H}_{0}(t)%
\mathbf{S}_{2}^{\dagger }=0,\text{ }\mathbf{S}_{2}\text{- definition}  \notag
\end{eqnarray}

By similar methods to the presented for the matrix $\mathbf{S}_{2}$, one can
derive the higher order correction i.e., find matrices $\mathbf{S}_{3},$ $%
\mathbf{S}_{4},..$ etc.

\section{Example}

\section{The case $N_{b}=2$}

Above, we described the method for deriving solution to the generalized MS
problem, where the evolution of a two-level quantum system is subjected to
pulsed external field and in the general case the ground and excited levels
have arbitrary level of degeneracy, $N_{a}$ and $N_{b}$ respectively. In
this section, we will illustrate these results with a specific example: when
the $\mathbf{b}$ set consists of two degenerate states, i.e. $N_{b}=2.$ This
case is interesting, because of the possible realizations of our model in
different real physical systems. Moreover, this special case has an exact
analytical solution, which is very rare for multidimensional systems.

\subsection{General case}

We label the ground states $\left\vert \psi _{i}\right\rangle $ $%
(i=1,2,...,N),$ where $N$ is arbitrary and we denote the two excited states $%
\left\vert \psi ^{\prime }\right\rangle $ and $\left\vert \psi ^{\prime
\prime }\right\rangle $. In this case the interaction operator $\mathbf{V}$ (%
\ref{V}) has $N$ columns and two rows, and its explicit form is given by

\begin{equation}
\mathbf{V}=\left[
\begin{array}{cc}
\Omega _{1}^{\prime } & \Omega _{1}^{\prime \prime } \\
\Omega _{2}^{\prime } & \Omega _{2}^{\prime \prime } \\
\vdots & \vdots \\
\Omega _{N}^{\prime } & \Omega _{N}^{\prime \prime }%
\end{array}%
\right] =\left[ \left\vert \Omega ^{\prime }\right\rangle ,\left\vert \Omega
^{\prime \prime }\right\rangle \right] ,
\end{equation}%
where $\left\vert \Omega ^{\prime }\right\rangle $ and $\left\vert \Omega
^{\prime \prime }\right\rangle $ are $N$-dimensional vectors comprising the
couplings between the ground states and the corresponding excited state%
\begin{equation}
\left\vert \Omega ^{\prime }\right\rangle =\left[
\begin{array}{c}
\Omega _{1}^{\prime } \\
\Omega _{2}^{\prime } \\
\vdots \\
\Omega _{N}^{\prime }%
\end{array}%
\right] ,~~~\left\vert \Omega ^{\prime \prime }\right\rangle =\left[
\begin{array}{c}
\Omega _{1}^{\prime \prime } \\
\Omega _{2}^{\prime \prime } \\
\vdots \\
\Omega _{N}^{\prime \prime }%
\end{array}%
\right] .
\end{equation}%
We write the product of $\mathbf{V}^{\dag }\mathbf{V}$%
\begin{equation}
\mathbf{V}^{\dag }\mathbf{V}=\left[
\begin{array}{cc}
\left\vert \Omega ^{\prime }\right\vert ^{2} & \left\langle \Omega ^{\prime
}|\Omega ^{\prime \prime }\right\rangle \\
\left\langle \Omega ^{\prime }|\Omega ^{\prime \prime }\right\rangle ^{\ast }
& \left\vert \Omega ^{\prime \prime }\right\vert ^{2}%
\end{array}%
\right] ,
\end{equation}%
with eigenvalues,
\begin{equation}
\lambda _{\pm }^{2}=\frac{1}{2}\left[ \left\vert \Omega ^{\prime
}\right\vert ^{2}+\left\vert \Omega ^{\prime \prime }\right\vert ^{2}\pm
\sqrt{D}\right] ,  \label{eigenvalues}
\end{equation}%
where $D$ denotes

\begin{equation}
D=\left( \left\vert \Omega ^{\prime }\right\vert ^{2}-\left\vert \Omega
^{\prime \prime }\right\vert ^{2}\right) ^{2}+4\left\vert \left\langle
\Omega ^{\prime }|\Omega ^{\prime \prime }\right\rangle \right\vert ^{2}=%
\frac{\left( \left\vert \Omega ^{\prime }\right\vert ^{2}-\left\vert \Omega
^{\prime \prime }\right\vert ^{2}\right) ^{2}}{\cos ^{2}2\theta },
\label{discriminant}
\end{equation}%
and we have introduced the following parameterizations,%
\begin{eqnarray}
\frac{2\left\vert \left\langle \Omega ^{\prime }|\Omega ^{\prime \prime
}\right\rangle \right\vert }{\left\vert \Omega ^{\prime \prime }\right\vert
^{2}-\left\vert \Omega ^{\prime }\right\vert ^{2}} &=&\tan 2\theta ,\quad
(0<2\theta <\pi ) \\
\arg \left\langle \Omega ^{\prime }|\Omega ^{\prime \prime }\right\rangle
&=&\sigma .
\end{eqnarray}

It is straightforward to obtain the eigenstates $|\widetilde{\psi }%
_{+}^{b}\rangle $ and $|\widetilde{\psi }_{-}^{b}\rangle $, corresponding to
$\lambda _{+}^{2}$ and $\lambda _{-}^{2},$ which represent the excited
states in the MS two- state systems,

\begin{subequations}
\label{b}
\begin{eqnarray}
|\widetilde{\psi }_{+}^{b}\rangle &=&\frac{1}{n_{+}}\left[
\begin{array}{c}
\left\langle \Omega ^{\prime }|\Omega ^{\prime \prime }\right\rangle \\
\lambda _{+}-\left\vert \Omega ^{\prime }\right\vert ^{2}%
\end{array}%
\right] =\left[
\begin{array}{c}
e^{i\sigma }\sin \theta \\
\cos \theta%
\end{array}%
\right] ,  \label{b+} \\
|\widetilde{\psi }_{-}^{b}\rangle &=&\frac{1}{n_{-}}\left[
\begin{array}{c}
\left\langle \Omega ^{\prime }|\Omega ^{\prime \prime }\right\rangle \\
\lambda _{-}-\left\vert \Omega ^{\prime }\right\vert ^{2}%
\end{array}%
\right] =\left[
\begin{array}{c}
e^{i\sigma }\cos \theta \\
-\sin \theta%
\end{array}%
\right] ,  \label{b-}
\end{eqnarray}%
with $n_{+}$ and $n_{-}$ normalization factors.

The next step is to find the Householder vectors $|\widetilde{\psi }%
_{+}^{a}\rangle $ and $|\widetilde{\psi }_{-}^{a}\rangle ,$ which are the
ground states in the MS\ two-state systems and define the propagator $%
\mathbf{U}_{\mathbf{a}}.$ They are the eigenstates of the $N$-dimensional
matrix,
\end{subequations}
\begin{equation}
\mathbf{VV}^{\dag }=\left\vert \Omega ^{\prime }\right\rangle \left\langle
\Omega ^{\prime }\right\vert +\left\vert \Omega ^{\prime \prime
}\right\rangle \left\langle \Omega ^{\prime \prime }\right\vert ,
\label{vv+}
\end{equation}%
corresponding to the same non-zero eigenvalues (\ref{eigenvalues}). We
construct them as superpositions of the interaction vectors with
coefficients $\alpha _{+}^{\prime },~\alpha _{+}^{\prime \prime },$ $\alpha
_{\_}^{\prime }$ and $\alpha _{\_}^{\prime \prime },$
\begin{equation}
\left\vert \widetilde{\psi }_{\pm }^{a}\right\rangle =\alpha _{\pm }^{\prime
}\left\vert \Omega ^{\prime }\right\rangle +\alpha _{\pm }^{\prime \prime
}\left\vert \Omega ^{\prime \prime }\right\rangle .
\end{equation}%
We determine these coefficients from the eigenvalue equations,
\begin{equation}
\mathbf{VV}^{\dag }(\alpha _{\pm }^{\prime }\left\vert \Omega ^{\prime
}\right\rangle +\alpha _{\pm }^{\prime \prime }\left\vert \Omega ^{\prime
\prime }\right\rangle )=\lambda _{\pm }(\alpha _{\pm }^{\prime }\left\vert
\Omega ^{\prime }\right\rangle +\alpha _{\pm }^{\prime \prime }\left\vert
\Omega ^{\prime \prime }\right\rangle ),
\end{equation}%
and from the normalization conditions,
\begin{equation}
\left\langle \widetilde{\psi }_{\pm }^{a}\right\vert \left. \widetilde{\psi }%
_{\pm }^{a}\right\rangle =1.
\end{equation}%
As a result we obtain%
\begin{eqnarray}
\left\vert \widetilde{\psi }_{+}^{a}\right\rangle &=&\frac{1}{\lambda _{+}}%
(\cos \theta \left\vert \Omega ^{\prime }\right\rangle +e^{-i\sigma }\sin
\theta \left\vert \Omega ^{\prime \prime }\right\rangle ),  \label{a+} \\
\left\vert \widetilde{\psi }_{-}^{a}\right\rangle &=&\frac{1}{\lambda _{-}}%
(\sin \theta \left\vert \Omega ^{\prime }\right\rangle -e^{-i\sigma }\cos
\theta \left\vert \Omega ^{\prime \prime }\right\rangle ).  \label{a-}
\end{eqnarray}

The results given in equations Eq.(\ref{b+}), Eq.(\ref{b-}), Eq.(\ref{a+})
and Eq.(\ref{a-}) define the block-structure of the $\mathbf{S}_{0}$ matrix
via Eq.(\ref{S-matrix-form}). Using Eq.(\ref{b+}), Eq.(\ref{b-}) the $%
\mathbf{B}_{0}$ matrix reads

\begin{equation}
\mathbf{B}_{0}=\left[
\begin{array}{cc}
e^{-i\sigma _{0}}\sin \theta _{0} & \cos \theta _{0} \\
e^{-i\sigma _{0}}\cos \theta _{0} & -\sin \theta _{0}%
\end{array}%
\right]  \label{B0}
\end{equation}

We will look for solution for the matrix $\mathbf{B}_{1},$ and without loss
of generality one can assume that $\mathbf{B}_{1}$ possesses tha same
functional form as $\mathbf{B}_{0}$

\begin{equation}
\mathbf{B}_{1}=\left[
\begin{array}{cc}
e^{-i\sigma _{1}}\sin \theta _{1} & \cos \theta _{1} \\
e^{-i\sigma _{1}}\cos \theta _{1} & -\sin \theta _{1}%
\end{array}%
\right] ,  \label{B1}
\end{equation}

where the matrix parameters $\sigma _{1}$ and $\theta _{1}$ need to be
derived. Using the matrix system of equations given by Eq.(\ref{S1-system})
the solutions for $\mathbf{A}_{1}$ and $\mathbf{B}_{1}$ can be derived.

\acknowledgments This work has been supported by the project QUANTNET -
European Reintegration Grant (ERG) - PERG07-GA-2010-268432.

\end{document}